\documentclass[12pt]{article}
\usepackage{graphics}
\usepackage{epsfig} 
\usepackage{hyperref}
\usepackage{graphics}
\usepackage{color}      
\usepackage{graphicx}

\title{\bf Thermally activated barrier crossing and stochastic resonance of a flexible  polymer chain in a piecewise linear bistable potential }
\author{Mesfin Asfaw \thanks{Present address: 
Research Institute for Electronic Science (RIES),
Hokkaido University, Japan; Electronic address:
taye@es.hokudai.ac.jp} \\
     Asian Pacific Center for Theoretical Physics,  Pohang 790-784,\\ South Korea } 
\date{Received: date / Revised version: date}

\begin{document}
\maketitle
\maketitle
\begin{abstract}

We study the stochastic resonance (SR) of a flexible  polymer
chain crossing over a piecewise linear bistable potential.
The dependence
of signal to  noise ratio  $SNR$  on  noise intensity $D$, coupling constant $k$  and  polymer length $N$ 
is studied via two state approximation.    We find that the response of signal to the
background noise strength is significant at  optimum values of $D_{opt}$, $k_{opt}$ and 
$N_{opt}$ which  suggests  novel means of  manipulating proteins or vesicles.  Furthermore, 
 the  thermally activated barrier crossing rate $r_{k}$  for the flexible  polymer
chain is studied. We find that 
    the crossing rate $r_{k}$ exhibits an optimal value at an optimal coupling constant $k_{opt}$; $k_{opt}$  decreases with $N$. As the chain  length $N$ increases, the escape rate for the center of mass $r_{k}$  monotonously decreases. On the other hand,   the crossing rate for the portion of polymer segment $r_s$ increases and saturates to a constant rate as   $N$ steps up.

\end{abstract}

\maketitle
 
\section{Introduction}

Understanding the physics of 
thermally activated barrier crossing for   systems  consisting of one or  more than one degree of freedoms \cite{c1,c2,c3,c4,c5,c6,c7,c8} is ubiquitous in many scientific disciplines. It has diverse physical applications and serves as a tool to understand stochastic paradigms such as stochastic resonance and thermal ratchets. Furthermore, for bistable systems, applying fields such as an external load or nonuniform temperature breaks their  symmetry  along the reaction coordinate  which may result in a unidirectional motion of particles.  
  In the last few decades several model systems, which are far from equilibrium,  have been proposed  \cite{c9, c10}. Especially when the external load (force) varies in time, coordination of the noise with time varying force may lead to the phenomenon of stochastic resonance  (SR) \cite{c11,c12} as long as the system is exposed to  weak sinusoidal signals.

The phenomenon of stochastic resonance (SR) is of interest  these days
because of its significant practical applications in a wide range of fields. SR depicts the phenomenon that systems, by utilizing the thermal background noise, enhance their performance when they are subjected  to
a small periodic signal. Since the pioneering work of Benzi. $et. al$ \cite{c11}, the idea of stochastic resonance has been broadened and implemented in many model systems \cite{c12, c13, c14, c15, c16, c17, c18, c19, c20, re20}. Recently the appearance of entropic stochastic resonance for a Brownian particle in a confined system has been reported in the work \cite{c21}. Unlike conventional energetic potential,
the effective potential of the confined systems may have an entropic nature due to the constrained regions. For tiny-scaled biological systems that frequently exist in a highly confined geometry, the entropic contribution to the effective potential is unavoidable and plays an important role in noise-induced resonant effects.

Often biological and soft matter systems such as polymers and membranes are complex and consist of several components. Their flexibility and connectivity lead to a fascinating dynamics under certain time varying external fields and noisy environment. The resonance behavior of these systems relies not only on the strength of the background temperature, but also on their size, flexibility and shape of the potential. Thus, the interplay between the shape of the potential, flexibility and size of polymer (protein) plays a crucial role in the enhancement of signal to noise ratio SNR or spectral amplification $\eta$ as reported in the works \cite{c15,c16,c17,c22}.

 Earlier,  Lindner $et$. $al$.  considered  linearly coupled damped bistable oscillators \cite{c15}. The dependence
of $SNR$ on the coupling strength $k$, number of oscillators $N$ and noise
intensity $D$ was envisioned numerically. It has been shown that the $SNR$ of
the oscillators depicts a global maximum ($SNR_{max}$) at certain $D$  and $k_{opt}$   for a
given  $N$. 
 Latter, utilizing $\Phi_{4}$ field theory, Marchesoni $et.al$ \cite{c16} independently checked the numerical results of Lindner $et$. $al$ \cite{c15} in large $N$ limit. Recently, Dikshtein $et.al$ \cite{c17} considered a polymer in a symmetric bistable potential where the two end points of the polymer are restricted by motionless pining points in the perpendicular direction to the symmetric potential. It has been shown that  the $SNR$ is enhanced for wider (less deep potential). More recently 
we studied  the stochastic resonance (SR) for a flexible polymer surmounting
a bistable  potential. Our analysis indicated that, 
due to the flexibility  that can
enhance crossing rate and change chain conformations at the barrier, 
the power amplification exhibits an optimal value   at optimal chain lengths $N_{opt}$
and elastic constants $k_{opt}$  as well as at  optimal noise strengths $D$ \cite{c22}.

The study of thermally activated escape rate of polymer has  been also the subject of many studies (see for example 
\cite{c29}). Since polymer is a flexible and an  extended object with a finite length, its rate or $SNR$ relies on    its coupling constant,  chain length, shape of the potential and   initial conformation along the reaction coordinate in a complicated manner. 
 Thus, more studies are needed to understand its complex dynamics. 
Most of the previous  studies  considered the center of mass motion.  However, in this paper  not only we examine the crossing rate for the center of mass motion $r_{k}$ but also the rate for  a portion  of the polymer segment $r_{s}$. We find that   $r_{k}<r_{s}$.

The aim of this paper is to explore the crossing rate and stochastic resonance of a flexible polymer chain in a piecewise linear bistable potential as a function of $k$, $N$ and  $D$ by considering initially  coiled chain conformation.
First we explore the escape rate of the chain as a function of the model parameters.
We show that $r_{k}$  monotonously decreases with  $N$; the rate $r_{s}$ increases and saturates to a constant value when $N$ steps up. Since the cooperation between the monomers increases with $k$, the rate increases as $k$ goes up. At certain $k_{opt}$,  $r_{k}$ attains an optimal value and  further increases in $k$ results in a lower rate as rigid polymer crosses the barrier at the expense of higher thermal kicks.

In this paper, 
  utilizing two state approximation, the dependence of $SNR$ on  $D$, $k$ and  $N$ is   examined. For globular polymer chain we show  the response of signal to the
background noise strength is significant at  optimum values of $D_{opt}$ and 
$N_{opt}$ which  suggests  novel means of  manipulating 
  (such as  efficient separation methods) not only for biopolymers, but also for proteins (vesicles) of different size.
 For the chain with a finite coupling constant $k$, the $SNR$ exhibits an optimal value at an optimal $k_{opt}$. The signal  to noise ratio for the center of mass motion  $SNR_{k}$ exhibits an optimal value at an optimum $N_{opt}$.  On the other hand, the $SNR_s$ for the polymer segment   monotonously increases with $N$. In addition, considering  temperature dependent coupling constant (entropic chain) where the elastic constant is given by $k=3D/l^{2}$,  
  we further confirm  that    $SNR$ for entropic chain shows  a broader  and  a higher peak than  a chain with temperature independent coupling constant.  Here, $l$ designates the Kuhn segment length.

  At this point  we stress that   even though the model and  its numerical approach 
 is completely different from our  previous work \cite{c22}, the results of this work  agree with that of the
previous work at least qualitatively.  One can note that
\begin{figure}[h]
\epsfig{file=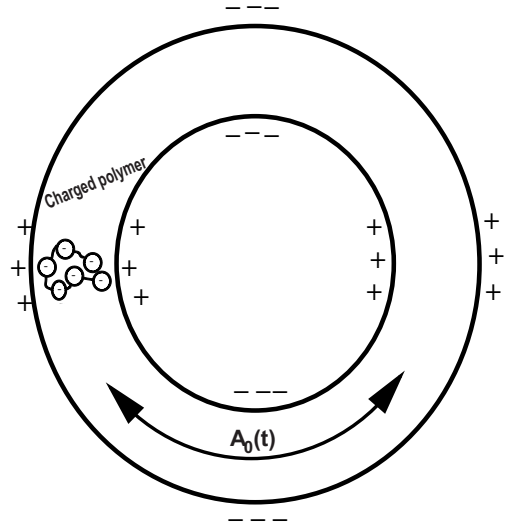,width=8cm}
\caption{Schematic diagram for initially coiled  polymer chain in a locally charged channel. The  fluidic channel is subjected  to an external
periodic red force $A_{0}(t)$ (AC field). }  
\end{figure}
the dynamics of such a system can be realized experimentally. One makes  negatively charged coiled  polymer chain, then put the polymer  within  positively and negatively  charged fluidic channel as shown in Fig. 1. The  fluidic channel is subjected  to an external
periodic force (AC field). Since the polymer is negatively charged, it encounters a difficulty of crossing through the negatively charged part of the  channel. Assisted by the thermal background kicks  along with its  conformational change, the polymer ultimately overcomes the barrier. The presence of time varying force,  may further enhance the rate of crossing.  One can sort or manipulate  polymer of specific coupling constant or chain length by tuning the angular frequency.

The rest of the  paper is organized as follows: in Section 2, we
present the model. In Section 3, we study the dependence
of the crossing rate  on the model parameters.
In section 4, we  discuss how the  $SNR$ for the flexible  chain behaves as a function of the model parameters.  
 Section 5 deals with summary and conclusion.

\section{The Model}

Let us  now  consider  a flexible polymer chain of size  $ N$ which  undergoes a Brownian motion in  one dimensional piecewise linear bistable potential as shown in Fig. 1. Considering only nearest-neighbor interaction between the polymer segments (the bead spring model), the Langevin equation that governs the 
 dynamics of the N beads ($n$=1,2,3 . . . $N$) in a highly viscous medium under the influence of   external potential $U(x)$ is given by  
\begin{equation}
\gamma{dx_{n}\over dt}=-k(2x_{n}-x_{n-1}-x_{n+1})-{\partial U(x_{n})\over \partial x_{n}}+  \xi_{n}(t)
\end{equation}
where the $k$ is the spring (elastic) constant of the chain while $\gamma$ denotes the friction coefficient.  $\xi_{n(t)}$ is assumed to be Gaussian  white noise satisfying
\begin{equation}
\left\langle  \xi_{n}(t) \right\rangle =0,~~~\left\langle \xi_{n}(t)  \xi_{n}(t+\tau) \right\rangle=2D\gamma
\delta(\tau)
\end{equation}
with $D= k_{B}T$ is the strength of the thermal noise. The external potential  each bead experiences is  considered to be a piecewise linear potential  \begin{equation} U(x)=\cases{
U_{0}[{-x\over L_{0}}-1],&if $ x \le -L_{0}$;\cr
U_{0}[{x\over L_{0}}+1],&if $-L_{0}\le x \le 0$;\cr
   U_{0}[{-x\over L_{0}}+1],&if $0 \le x \le L_{0};$\cr
   U_{0}[{x\over L_{0}}-1],&if $ x \ge L_{0};$\cr
   }\end{equation} where $U_{0}$ and $2L_{0}$ denote the barrier height and the width of the piecewise linear bistable potential, respectively. 
   If one considers only   the center of mass motion,  the second term in Eq. (1) vanishes.  For    a globular polymer
chain, where the
coupling (spring) constant $k$ is infinity,   the Langevin equation (Eq. (1)) for the center of mass (cm) takes a simple form
\begin{equation} 
N\gamma \frac{dx_{cm}}{dt}=\cases{N\frac{U_{0}}{L_{0}}  +
        \xi(t),&if $x \le -L_{0}$;\cr
        -N\frac{U_{0}}{L_{0}}  +
        \xi(t),&if $-L_{0}< x \le 0$;\cr
N\frac{U_{0}}{L_{0}}  +
        \xi(t),&if $0< x \le L_{0}$;\cr
  -N\frac{U_{0}}{L_{0}}  +
        \xi(t),&if $x > L_{0}.$\cr
   }\end{equation} 
   \begin{figure}[ht]
\epsfig{file=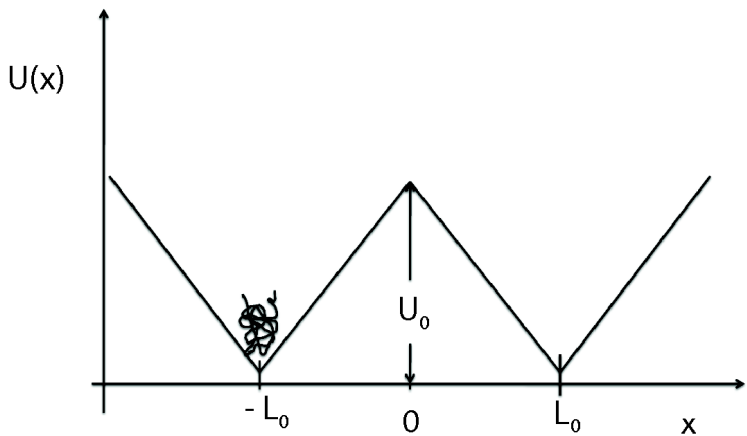,width=10cm}
\caption{Schematic diagram for initially coiled  polymer chain in a piecewise linear bistable potential. The potential wells and the barrier top are located at $x=\pm L_{0}$ and  $x=0$, respectively. Due to the thermal background kicks, the polymer ultimately crosses the barrier assisted by its conformational change along the reaction coordinate.  }  
\end{figure}

   \section{Polymer escape rate}

We consider a polymer which has initially coiled conformation along  the  piecewise linear bistable potential as shown in Fig. 2.  Due to the thermal background kicks, the polymer presumably crosses the barrier. The crossing rate of the chain strictly relies on the chain  length,  coupling constant, barrier height and noise strength.
For a compact polymer   ($k \to \infty$), the  monomers feel the same force along the reaction coordinate (see Eq. (4)) and due to  lack of chain conformational change at the top of the barrier, its escape rate is considerably lower than a polymer of  finite $k$ value. On the other hand, the jumping rate also depends on the choice of coordinate system. The rate for  the center of mass motion is a decreasing function of $N$ while the escape rate for the portion of polymer segment increases with $N$. The thermally activated rate strictly relies   on the chain initial conformation along the piecewise linear bistable potential  for the case where the microscopic relaxation
time of the polymer is significant compared to the crossing time.  In this case,
initially stretched polymer crosses the barrier faster than coiled chain. This is because coiled polymer first stretches before crossing the barrier.   The degree of stretching depends on the relaxation time of the polymer which itself relies on the chain length, coupling constant and the thermal background kicks. For short polymer chain  surmounting a potential barrier that is large compared to $D$, $NU_{B}\gg D$,  the chain crossing time is considerably larger than  its  relaxation time  and  hence the crossing time may be independent of chains initial conformation.
The activated barrier crossing considered in this work 
strictly applicable only to the case where the crossing time
is much larger than any microscopic relaxation
times of the polymer.

Before we discuss how the flexible  polymer  chain in  the double-well potential behaves, let us first calculate the crossing rate for  a globular polymer chain.
The problem of surmounting a piecewise linear bistable potential for a single Brownian particle in high friction limit was  addressed in the work \cite{c24}. Following  the same approach,
the mean first passage time MFPT  for  the compact polymer crossing  over a high potential barrier    is given by 
        \begin{equation}\label{eq:8}
        MFPT = \frac{\gamma}{ D}\left(\frac{DL_{0}}{NU_{0}}\right)^2 e^{\frac{NU_{0}}{D}}
 \end{equation}
               while the crossing rate  
               \begin{equation}\label{eq:8}
                r_{k}={D\over \gamma}\left(\frac{NU_{0}}{DL_{0}}\right)^2 e^{\frac{-NU_{0}}{D}}
                  \end{equation}  
                  is the inverse of MFPT.
                  The dependence of the crossing rate $r_{k}$  or equivalently
the $MFPT$  on the potential height $U_{0}$, width of the potential $2L_{0}$ and chain length $N$ can be analyzed
by exploiting  Eq. (6). When $U_{0}$ increases, the polymer encounters
a difficulty of jumping the piecewise linear bistable potential and as a result 
  $r_{k}$ declines.   When  $N$ increases,  $r_{k}$ decreases as large polymer crosses the potential barrier at the expense of higher thermal kicks. On the other hand,   as $L_{0}$  steps up, the $MFPT$ for the polymer to reach to the other side of the well increases which implies $r_{k}$ monotonously decreases.

For the chain with  a finite coupling constant,  we analyze  the crossing rate  via numerical simulation. We introduce dimensionless parameters: ${\bar x}=x/L_{0}$, $\tau=\gamma L^{2}/U_{0}$, ${\bar k}=kL^{2}/U_{0}$ and  ${\bar t}=t/\tau$. Hereafter all the quantities are rescaled (dimensionless) so the bars will be dropped.  The behavior of the system is analyzed  by  integrating the  Langevin equation (1)   (employing Brownian dynamics simulation). In the simulation, coiled  polymer chain  with  $N$ monomers is initially  situated in one of the potential  wells. Then the trajectories for the center of mass of the polymer or the portion of polymer segment is simulated by considering  different time steps $\Delta t$ and time length $t_{max}$. In order to ensure the numerical accuracy, up to $5 \times  10^{5}$ ensemble averages have been obtained.
\begin{figure}[h]
\epsfig{file=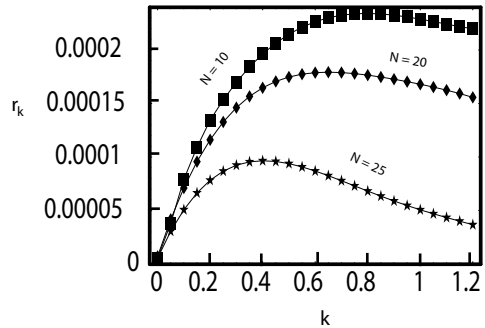,width=10cm}
\caption{Polymers crossing rate $r_{k}$  as a function of coupling constant $k$ for different chain length $N$ and   noise strength $D=0.5$. The simulation results are obtained directly by integrating Eq. (1). The figure exhibits that the rate increases as $N$ declines and attains an optimal value at an optimal elastic constant $k_{opt}$. The optimal coupling constant shifts to the right as $N$ decreases. }  
\end{figure}
\begin{figure}[ht]
\centering
{
    \includegraphics[width=6cm]{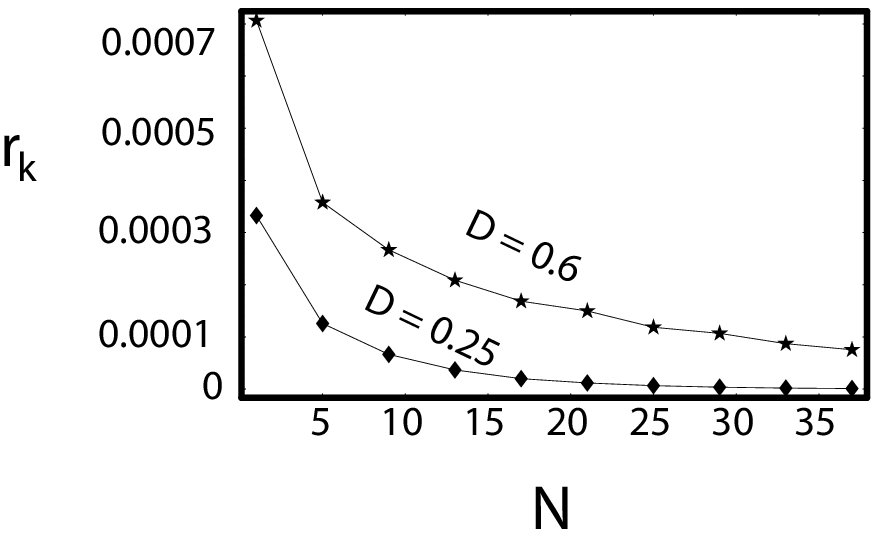}}
\hspace{1cm}
{
    \includegraphics[width=6cm]{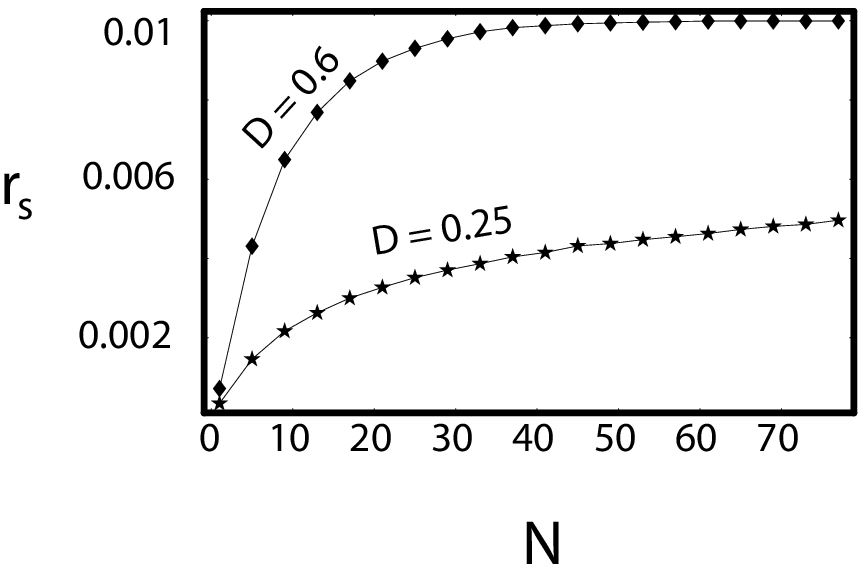}
}
\caption{(a) Dimensionless rate $r_{k}$ versus  $N$. (b) The escape rate for polymer segment $r_{s} $  versus  $N$. The data are obtained numerically for different rescaled noise strength $D$ and  coupling constant $k=0.5$.} 
\label{fig:sub} 
\end{figure}

 Let us now vary the rescaled spring
constant $k$ of the chain. 
Figure 3 plots the dependence of $r_{k}$ on the rescaled $k$ for parameter choice $N=10$, $N=20$, $N=25$ and $D=0.5$.  For small $k$, since the monomers become non-interacting (non-cooperative),  $r_{k}$  tends to be smaller.  When coupling strength between the monomers $k$ further increases, the tendency for interconnected monomers to assist each other increases   
 as a result $r_{k}$ steps up. At  certain optimal $k_{opt}$, $r_{k}$ attains
an optimum value.  Further increasing in $k$ results  in a lower  crossing rate as larger thermal energy
is required to drive the rigid chain across the reaction coordinate. The same figure depicts that  $k_{opt}$ is a decreasing  function of $N$.  
The coupling constant $k_{opt}$ relies on rescaled noise strength $D$ in a manner $k=3DL^{2}/l^{2}$.  The crossing rate for this entropic chain has been analyzed and compared with the  chain of  a finite $k$. The numerical analysis shows  the rate for entropic chain  is  considerably higher; further 
 details  will be reported elsewhere.

 Figure 4a  depicts the plot of $r_{k}$  as a function of $N$ for parameter choice $D=0.25$,  $D=0.6$ and  $k=0.5$.  The crossing rate for the center of mass motion
 monotonously decreases with $N$. On the contrary, the rate of the polymer segment $r_{s}$  increases and saturates to a constant value as $N$  and $D$ step up as shown in Fig. 4b.

\section{Stochastic resonance }

In the presence of time varying signal,  the interplay between noise, sinusoidal driving force together with  chain flexibility and  chain length
in the bistable system may lead the system into stochastic resonance provided
 the random tracks are adjusted in  an optimal way to the recurring external
force.  Next  we study the dependence of the SR on the model parameters employing two state approximations 
without considering a continuous diffusion dynamics.

In the presence of  a periodic signal $A_{0}\cos{(\Omega t)}$,  
the Langevin equation that governs the  dynamics of the system  is given by 
\begin{equation}
\gamma{dx_{n}\over dt}=-k(2x_{n}-x_{n-1}-x_{n+1})-{\partial U(x_{n})\over \partial x_{n}}+ A_{0}\\cos{(\Omega t)}+ \xi_{n}(t)
\end{equation}
where $A_{0}$ and $\Omega$  are 
the amplitude and angular frequency, respectively.
\begin{figure}[h]
\epsfig{file=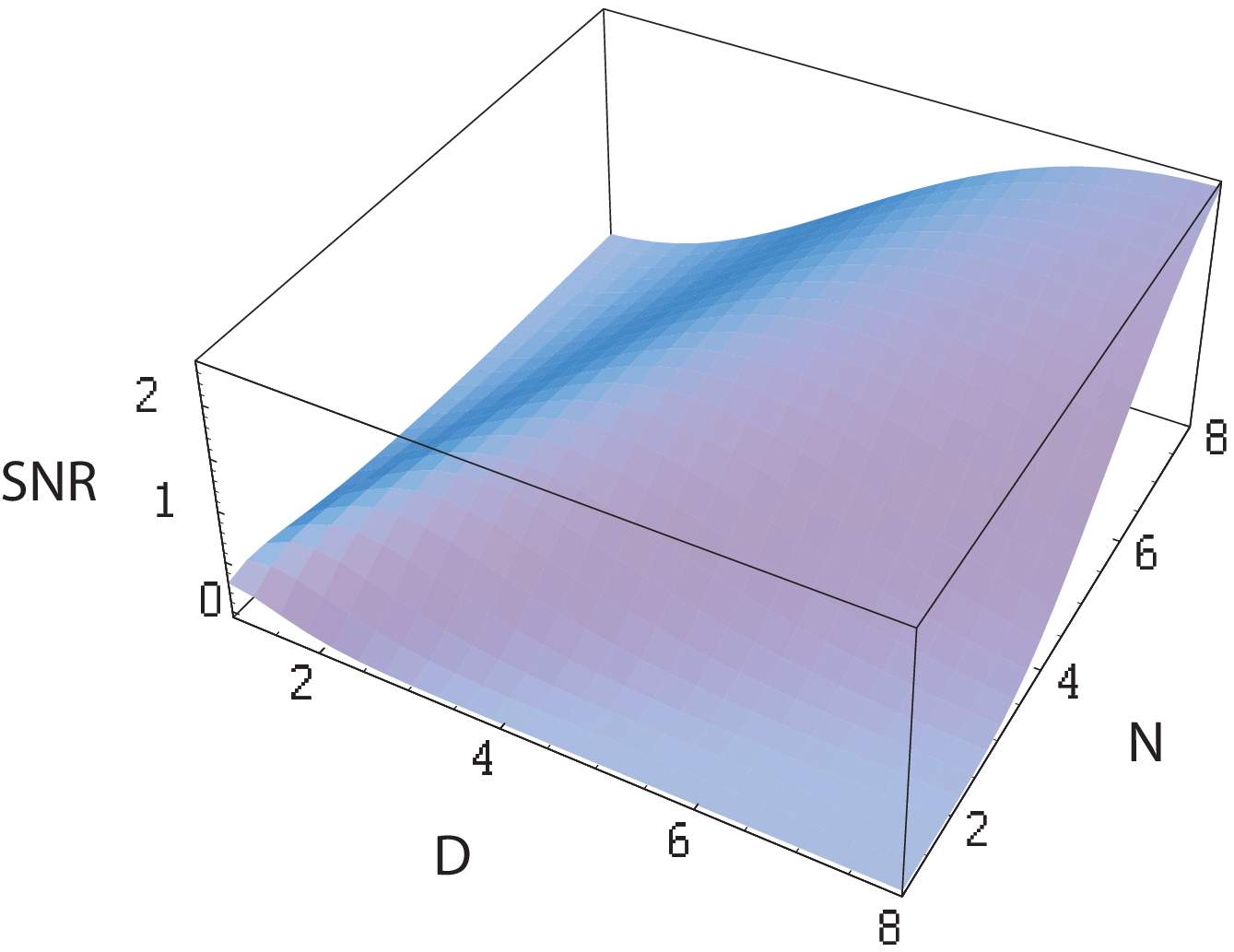,width=8cm}
\caption {(Color online)The dependence of $SNR \times 10^{-6}$ (for compact polymer chain)  on  $N$ and  $D$ for parameter choice $A_{0}=0.05$.  The $SNR$ is obtained via  Eqs. (6) and (10).     }  
\end{figure}

\begin{figure}[h]
\epsfig{file=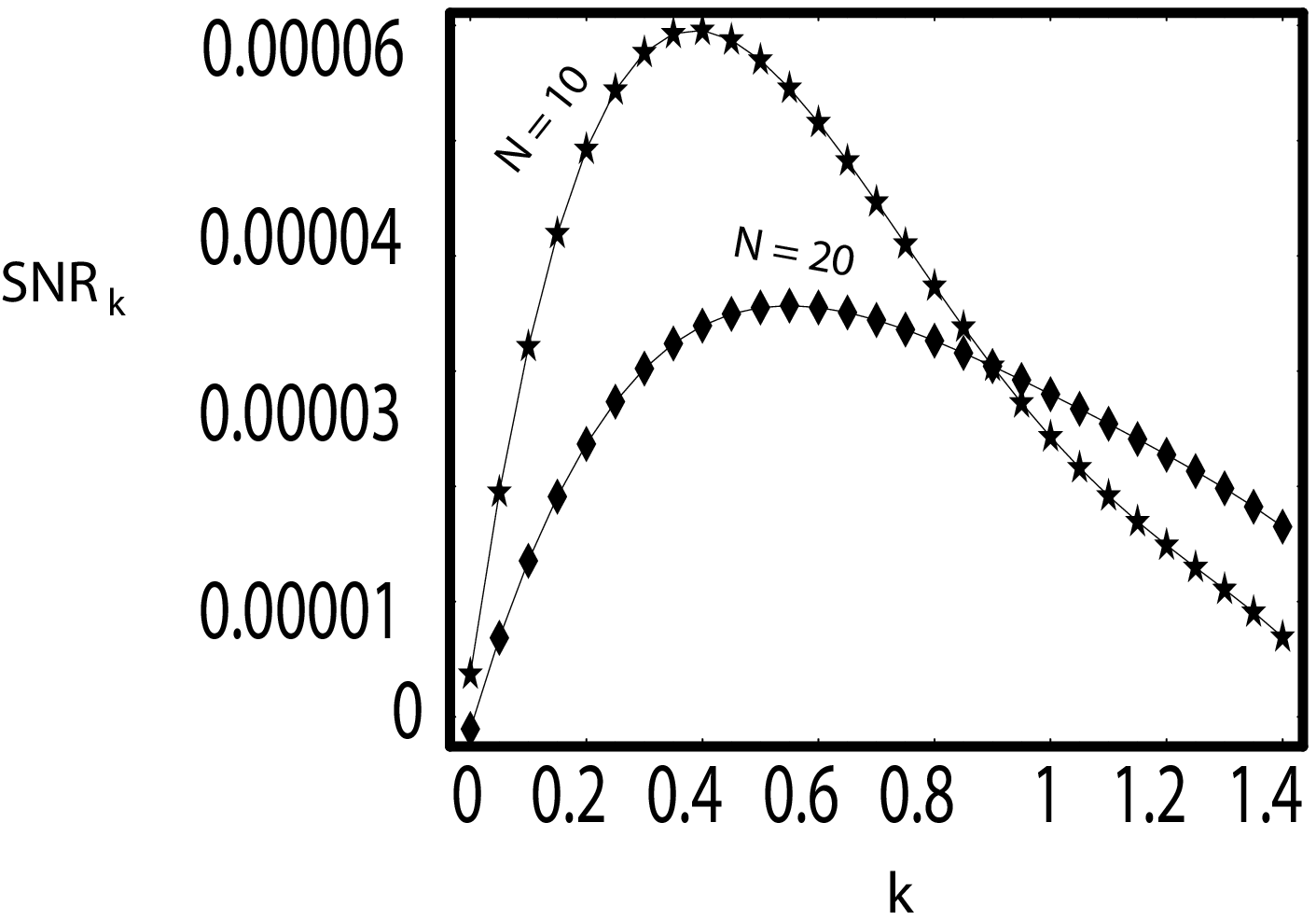,width=8cm}
\caption{The dependence of $SNR_k$ as a function of coupling constant $k$ for different chain length $N$, for noise strength $D=0.25$ and  amplitude $A_{0}=0.1$. The simulation results are obtained directly by integrating Eq. (1). The figure exhibits that the $SNR$ increases as $N$ declines and attains an optimal value at an optimal elastic constant $k_{opt}$. Unlike the corresponding rate, the optimal coupling constant shifts to the left as $N$ decreases.}  
\end{figure}

Employing two state model approach \cite{c12,c23}, two discrete states $x(t) = \pm L_0$ are
considered. Let us denote $n_+ $ and $n_-$ to be the probability  to find  the
polymer segment in the right ($L_{0}$) and  in the left ($-L_{0}$) sides of the 
potential wells, respectively. In the presence time varying  signal, the
master equation that governs  the time evolution of $n_{\pm}$ is given by
\begin{equation}
{\dot n_{\pm}(t)}=-W_{\pm}(t)n_{\pm}+W_{\mp}(t)n_{\mp}
\end{equation}
where $W_{+}(t)$ and $W_{-}(t)$ corresponds to the time dependent transition probability
 towards the right ($L_{0}$) and  the left ($-L_{0}$) sides of the 
potential wells. The time dependant rate \cite{c12,c23} takes a simple form
\begin{equation}
W_{\pm}=r_{k} exp\left[\pm {L_0 N\over U_{0}D}A_{0}\cos{(\Omega t)}\right]
\end{equation}
where $r_k$ is the Kramers rate for  the polymer in the absence of periodic force
$A_0 = 0$. We consider the case
 where $N \gg D$. For sufficiently small amplitude, one finds
the signal to noise ratio
\begin{equation} 
SNR=\pi r_{k}({A_{0}N L_0\over U_{0}D})^2.
\end{equation}

\begin{figure}[ht]
\centering
{
    \includegraphics[width=6cm]{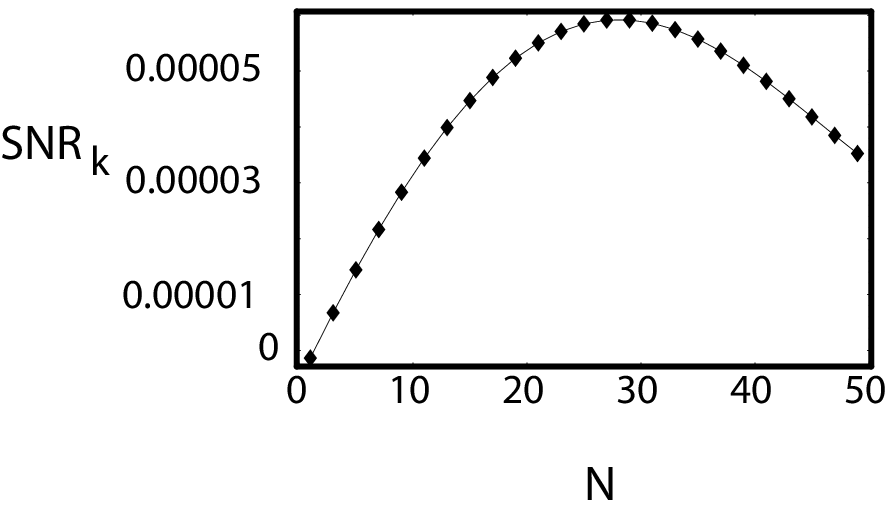}}
\hspace{1cm}
{
    \includegraphics[width=6cm]{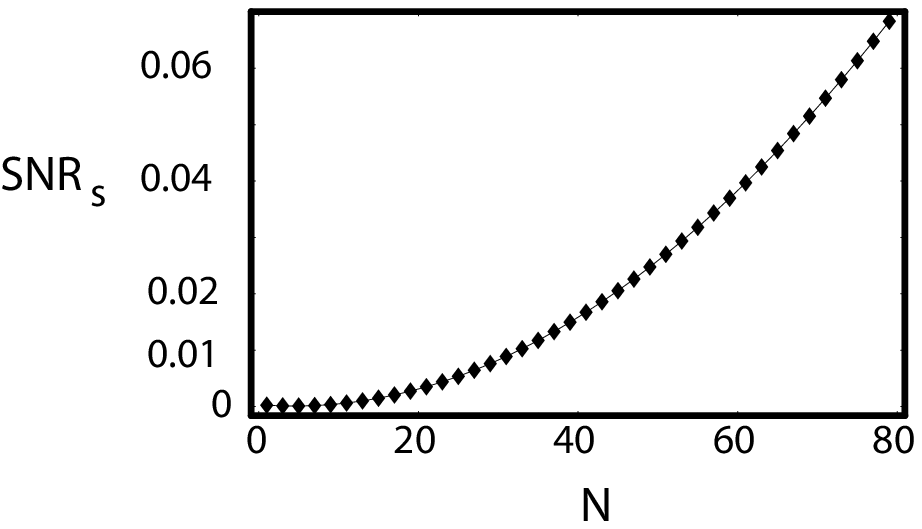}
}
\caption{(a)  $SNR_{k}$ versus  $N$. (b)  $SNR_{s}$ for polymer segment as a function of   $N$. The data are  obtained numerically for noise strength $D=0.5$, amplitude $A_{0}=0.1$ and  coupling constant $k=0.5$.} 
\label{fig:sub} 
\end{figure}

Before we explore how the  $SNR$ for globular polymer  behaves   as a function of $N$, we introduce additional  dimensionless parameter:  ${\bar A_{0}}=A_{0}L/ U_{0}$. From now on   for convenience   the bar will be dropped. 
Figure 5     depicts the plot for the  $SNR$ for a globular polymer  chain versus    $D$ and $N$  for fixed  values of the driving frequency and  potential width. The $SNR$ exhibits   nonmontonous noise strength dependence revealing  a peak at an optimal noise strength $D_{opt}$.  
$D_{opt}$  steps up as $N$  inclines. This is because longer polymer crosses the barrier at the expense of higher thermal excitation. 
  The same figure exhibits that  the $SNR$  peaks at optimum  chain length $N_{opt}$. This suggests  novel means of  manipulating 
   not only for biopolymers, but also for proteins (vesicles) of different size. One can note that since compact polymer lacks the flexibility to conform to the piecewise linear bistable potential,  the peak of the   $SNR$ is less than that of a flexible polymer chain.
  
  For the chain with a finite coupling constant $k$, the resonance behavior of the system is examined numerically. Figure 6 shows  
   the  plot for the $SNR_k$ as  a function of $k$ for  different values of $N$. Other parameters are  fixed as   $D=0.25$ and $A=0.1$.  For small $k$, the monomers tend to be non-interacting as a result the  $SNR$ becomes very small.  
   The $SNR$ peaks at an optimal chain length $k_{opt}$ and further increasing in $k$ leads again to a smaller $SNR$ as the rigid polymer is not flexible enough to adjust itself with the time varying force.  The optimal spring constant $k_{opt}$ increases as $N$ increases.
   
   Figure 7a  plots the rate $SNR_{k}$ versus $N$.  We observe an increase in the signal to noise ratio when the number of monomers increases. The $SNR$ peaks at an optimal chain length $N_{opt}$ and declines  again  as the chain length increases. On the contrary, $SNR_{s}$ monotonously increases with $N$ (see Fig. 7b).  It is worth noting that since the flexible polymer chain responds in more cooperative and coherent manner to the time varying force, it  exhibits a higher resonance peak than a globular polymer.

   Finally we emphasize   that even though the model and  its numerical approach 
 is completely different from  our  previous work \cite{c22}, the results  agree with that of the
previous work at least qualitatively.   For instance, similar to this work, in  our previous  work we found that the response of signal to the
background noise strength is significant at  optimum values of $D_{opt}$, $K_{opt}$ and 
$N_{opt}$. Our previous work \cite{c22} dealt with the study of SR for a flexible polymer chain on Kramer's potential. Depending on the chain length $N$ and spring constant $k$, the chain takes  
either   coiled or  stretched conformation at the top of the barrier. When the chain is either in coiled or stretched state, the resonance is much larger than globular state revealing  the  intrinsic flexibility  of the chain facilitating faster crossing.  In the globular limit $k\to \infty$, the resonance becomes small as the compact polymer lacks the flexibility to conform to the driving force. Furthermore, we showed that the power amplification peaks at an  optimal chain length
and elastic constant as well as at  an optimal noise strength.

\section{Summary and conclusion}

 In summary, in this work we  explore  the crossing rate and stochastic resonance of a flexible polymer chain in a piecewise linear bistable potential.  We investigate  the chain's  escape rate  as a function of different  model parameters.
  $r_{k}$  monotonously decreases with  $N$ while  the rate $r_{s}$ increases and saturates to a constant rate when $N$ steps up. Due to lack of cooperation between the monomers, the rate is considerably  small for a smaller $k$.  The  crossing rate increases as $k$ goes up. At certain $k_{opt}$,  $r_{k}$ attains an optimal value. Further increasing in  $k$ results in a lower rate as rigid polymer crosses the barrier at the expense of higher thermal kicks.

Employing  two state approximation, the dependence of $SNR$ as a function of model parameters  is  studied. For a globular polymer chain we show  the response of signal to the
background noise strength is significant at  optimum values of $D_{opt}$ and 
$N_{opt}$ which  suggests  novel means of  manipulating 
   proteins (vesicles) of different size. On the other hand, 
 for the chain with a finite coupling constant $k$, the $SNR$ exhibits an optimal value at an optimal $k_{opt}$. The signal  to noise ratio for the center of mass $SNR_{k}$ exhibits an optimal value at an optimum $N_{opt}$. The SNR for the polymer segment $SNR_{s}$  monotonously increases with $N$. 
  
 In  conclusion,  in this work,  by  introducing a different  model system than the previous work \cite{c22},  we recapture the previous  results  at least qualitatively. Not only we assess the resonance property of the system, but we also further   explore the barrier crossing rate by varying different model parameters. Since polymers are interconnected  and flexible systems,  they exhibit interesting cooperative dynamics when they are exposed to time varying external fields and noises. Understanding  their dynamics is crucial  not only for  novel means of manipulating proteins such as  DNA or RNA molecules in a nanofludic or microfludic channels,  but also   to understand  how such  systems self-organize their flexible degrees of freedom. 
Thus, this theoretical study is crucial not only
for the fundamental understanding of polymer physics,
but also  provides a basic paradigm 
in which to understand the self-organization and cooperativity induced by the chain 
flexibility and fluctuations.

\section{Acknowledgment}

I acknowledge the support of APCTP. I would like to thank Prof. W. Sung for the interesting discussions I had during my  visit  at APCTP, Korea.   I would like also to
thank Prof. Mulugeta Bekele  for his helpful comments, suggestions and critical
reading of this manuscript.


\begin{thebibliography}{47}
\bibitem{c1} H.A. Kramer. Physica {\bf 7}, 284 (1940).
\bibitem{c2} P. H\"anggi, P. Talkner and M. Borkovec, Rev. Mod. Phys. {\bf 62}, 251 (1990).
\bibitem{c3} P.J. Park and W. Sung, J. Chem. Phys. {\bf 111}, 5259 (1999). 
\bibitem{c4} S. Lee and W. Sung, Phys. Rev. E {\bf 63}, 021115 (2001).
\bibitem{c5} P. H\"anggi,  F.  Marchesoni and P.  Sodano,  Phys. Rev. Lett.  {\bf 60},  2563  (1988).      
\bibitem{c6}  F. Marchesoni, C. Cattuto and G. Costantini, Phys. Rev. B, {\bf  57}, 7930 (1998). 
\bibitem{c7}   P. H\"anggi  and F. Marchesoni, Rev. Mod. Phys. {\bf 81}, 387 (2009).
\bibitem{c8} K.L. Sebastian and Alok K.R. Paul, Phys. Rev. E {\bf 62}, 927 (2000).
\bibitem{c9} M. Asfaw and M.  Bekele, Eur. Phys. J. B {\bf 38}, 457 (2004).
\bibitem{c10} M. Asfaw, Eur. Phys. J. B {\bf 65}, 109 (2008).
\bibitem{c11} R. Benzi, G. Parisi, A. Sutera and A. Vulpiani, Tellus  {\bf 34}, 10 (1982).
\bibitem{c12} L.  Gammaitoni, P. H\"anggi, P. Jung and F. Marchesoni, Rev. Mod. Phys. {\bf 70}, 223 (1998).
\bibitem{c13} A.  Neiman and W. Sung, Phys. Lett. A {\bf 223}, 341 (1996).
\bibitem{c14} P. Jung, U. Behn, E. Pantazelou, and F. Moss, Phys. Rev. A {\bf 46}, R1709
(1992).
\bibitem{c15}  J. F. Lindner, B. K. Meadows, W. L. Ditto, M. E. Inchiosa, and A. R. Bulsara, Phys. Rev. Lett. {\bf 75}, 3 (1995); Phys. Rev. E {\bf 53}, 2081 (1996).
\bibitem{c16} F. Marchesoni, L. Gammaitoni, and A. R. Bulsara, Phys. Rev. Lett. {\bf 76}, 2609 (1996). 
\bibitem{c17} Igor E. Dikshtein, Dmitri V. Kuznetsov and Lutz Schimansky-Geier,  Phys. Rev. E. {\bf 65}, 061101 (1996). 
\bibitem{c18} I. Goychuk and P. Hanggi, Phys. Rev. Lett. {\bf 91}, 070601 (2003).
\bibitem{c19} H. Yasuda et al., Phys. Rev. Lett. {\bf 100}, 118103 (2008).
\bibitem{c20} J. M. G. Vilar and J. M. Rubi, Phys. Rev. Lett. {\bf 78}, 2886 (1997).
\bibitem{re20} J. F. Lindner, M. Bennett, and K. Wiesenfeld, Phys. Rev. E {\bf 73}, 031107 (2006).
\bibitem{c21} P. S. Burada, G. Schmid, D. Reguera, M. H. Vainstein, J. M. Rubi, and P. H\"anggi, Phys. Rev. Lett. {\bf 101}, 130602 (2008).
\bibitem{c22} M.  Asfaw and W. Sung, manuscript accepted for publication in Euro. Phys. Lett.
\bibitem{c29} K. L. Sebastian and A. K. R. Paul, Phys. Rev. E {\bf 62}, 927 (2000).
\bibitem{c23} B. McNamara, K. Wiesenfeld, Phy. Rev. A, {\bf 39}, 4854 (1989).
\bibitem{c24}Z. Getahun1, M. Asfaw and M. Bekele1, Manuscript submitted to JIMPB.


\end{thebibliography}
\end{document}